\documentstyle [12pt]{article}
\makeatletter \oddsidemargin  0in \evensidemargin 0in
\textwidth16cm \RequirePackage[dvips]{graphicx} \textheight 20.5cm
\newcommand{\singlespacing}{\let\CS=\@currsize\renewcommand{\baselinestretch}{1.5}\tiny\CS}
\makeatletter \oddsidemargin  -.1in \evensidemargin -.1in
\newcommand{\doublespacing}{\let\CS=\@currsize\renewcommand{\baselinestretch}{1.35}\tiny\CS}

\def\@citex[#1]#2{\if@filesw\immediate\write\@auxout{\string\citation{#2}}\fi
  \def\@citea{}\@cite{\@for\@citeb:=#2\do
    {\@citea\def\@citea{,\linebreak[0]\hskip0pt plus .2em}%
      \@ifundefined{b@\@citeb}%
    {{\bf ?}\@warning{Citation `\@citeb' on page \thepage\space undefined}}%
      \hbox{\csname b@\@citeb\endcsname}}}{#1}}

\newtheorem{rule-def}[theorem]{Rule}
\begin{document}
\title{\bf Entanglement and Mixed ness of Locally Cloned Non - Maximal W - State }
\author{Indranil Chakrabarty $^{1,3}$\thanks{Corresponding author:
E-Mail-indranilc@indiainfo.com }, Sovik Roy $^{2,3}$,\\ Nirman Ganguly $^{1,3}$,Binayak S. Choudhury$^{3}$ \\
$^1$Heritage Institute of Technology, Kolkata - 107, WestBengal, India.\\
$^2$Techno India,Salt Lake, Kolkata - 91, West Bengal, India\\
$^3$Bengal Engineering and Science University, Shibpur, West
Bengal, India }
\date{}
\maketitle{}
\begin{abstract}
In this work we describe a protocol by which two of three parties generate  two bipartite  entangled state among
themselves without involving third party, from a non maximal W state or W - type state
$|X\rangle=\alpha|001\rangle_{123}+\beta|010\rangle_{123}+\gamma|100\rangle_{123}, \alpha^{2} + \beta^{2} +
\gamma^{2} = 1$ shared by three distant partners. Also we have considered the case $\beta=\gamma$, to obtain a range
for $\alpha^2$, for which the local output states are separable and non local output states are inseparable. We
also find out the dependence of the mixed ness of inseparable states with their amount of inseparability, for that
range of $\alpha^2$.
\end{abstract}
\section{Introduction: }For decades , quantum entanglement have
been the focus of much of the work in the foundation of quantum
mechanics. In particular, it's genesis comes with the concepts of
non - separability, the violation of Bell Inequalities and EPR
paradox. Creation and operation with entangled states are
essential for quantum information application.Some of the
applications are quantum teleportation [1], quantum dense coding
[2], quantum error correction [3], quantum cryptography [4]. Hence
quantum entanglement has been viewed as an essential resource for
quantum information processing and all of these applications
depend upon the strength of quantum entanglement. One of the most
important aspects of quantum information processing is that
information can be 'encoded' in non - local correlations
(entanglement) between two
separated particles.\\\\
The present work deals with the local copying and partial
broadcasting of entangled pairs [5]. So we should have a
preliminary idea about what we actually mean by partial
broadcasting.\\\\
Let Alice and Bob share an inseparable (entangled) state whose
density operator is given by $\rho_{AB}^{id}$.We will use two
quantum copiers $Q_{1}$ and $Q_{2}$ (the density operator
$\rho_{Q_{1}Q_{2}}$ describing the input state of two quantum
copiers is separable) to locally copy A and B such that at the
output level of A and B two other states C and D are produced
respectively. As a result of this copying, we obtain, out of two
systems A and B, four other systems described by a density
operator $\rho_{ABCD}^{out}$. Now if we can see that the states
$\rho_{AD}^{out}$ and $\rho_{BC}^{out}$ are inseparable while the
states $\rho_{AC}^{out}$ and $\rho_{BD}^{out}$, which are produced
locally, are separable, then we can say that we have partially
broadcasted (cloned, split) the entanglement (inseparability) that
was present in the input state.\\\\
Our motivation for this present work is basically  two fold
:\\
(1) We discussed a protocol by which we can generate two
bipartite entangled states between two parties Alice and Bob from
a W-type state shared by Alice, Bob and Charlie without involving
Charlie at all.For doing so, we have started with a three party
entangled state and by Buzek - Hillery quantum cloners [6], we
have locally copied the first two particles.Then without losing
generality we have traced out the third party as well as the
machine state, to produce four bipartite local and non - local
output state . Then we have considered the case $\beta=\gamma$ to
find out the range of $\alpha^2$ for which local output states are
separable and the non local output states are inseparable.A
pragmatic inference, if we can find the output states to be
entangled , is that we can use them as channels to encode more
information. Also it is seen that the third party in the three
party entangled pair has no effect on the whole process.\\
(2) Keeping the range of $\alpha^2$ fixed we find out that whether
the mixed ness of the partially broadcasted subsystems obtained
from the entangled input system does have any effect on their
entanglement (inseparabilty) i.e. whether or not the amount of
their entanglement have any relation with mixed ness. When
entangled , we find out the amount of entanglement and check if
there exists any relation between the
amount of entanglement and mixed ness of the output states.\\
The paper can thus be summarized as
follows:\\\\
In section 2, we have taken a non - maximal W - state (or W - type
state) which is defined as $|\psi\rangle = \alpha|001\rangle_{123}
+ \beta|010\rangle_{123} + \gamma|100\rangle_{123}, \alpha^{2} +
\beta^{2} + \gamma^{2} = 1$. Then we locally copy first two qubits
and then we trace out third qubit and machine state to obtain two
bipartite entangled state . The logic behind taking W - type state
and not W - state ($\alpha = \beta = \gamma = \frac{1}{\sqrt{3}})$
is that in case of W - state all the local and non - local
bipartite output states become separable and there
is nothing to prove further.\\
In section 3, by taking $\beta=\gamma$ we obtain the range of
$\alpha^2$ for which the local output states are separable and non
local output states are inseparable. \\
In section 4, we made a comparative study of the mixed ness and
amount of entanglement of the entangled and non - entangled
bipartite states with the help of 'Linear Entropy'
and 'Concurrence'.\\
In section 5, i.e. 'Conclusion', we have reviewed the previous
sections and have given our concluding remark.\\\\
\section{Analysis of non-maximal W state (or W type state):}
For this we shall start with W type state which is given by-
\begin{eqnarray}
|X\rangle=\alpha|001\rangle_{123}+\beta|010\rangle_{123}+\gamma|100\rangle_{123}
,where,\alpha^{2}+\beta^{2}+\gamma^{2}=1
\end{eqnarray}. Here we do not consider the cases where, $\alpha =
1, \beta = 0, \gamma = 0$ or $\alpha =0, \beta = 0, \gamma = 1$ or
$\alpha = 0, \beta = 1, \gamma = 0$. We shall clone the first two
bits of state $ (1) $ with the help of Buzek-Hillery quantum
cloning machine, where transformations are given by the following:
\begin{eqnarray}
\ |0\rangle\rightarrow\sqrt{\frac{2}{3}}|00\rangle|\uparrow\rangle
+ \sqrt{\frac{1}{6}}[|01\rangle + |10\rangle]|\downarrow\rangle
\end{eqnarray}
\begin{eqnarray}
|1\rangle\rightarrow\sqrt{\frac{2}{3}}|11\rangle|\downarrow\rangle
  +\sqrt{\frac{1}{6}}[|01\rangle+|10\rangle]|\uparrow\rangle
\end{eqnarray}

Here $ |\uparrow\rangle$ and $ |\downarrow\rangle$ are machine
states.After cloning the first two bits and tracing out the
machine states as well as the third bit we get the density matrix
of the resulting state as follows:
\begin{eqnarray}
\rho_{1245}=\alpha^{2}[\{\frac{2}{3}|00\rangle_{14}\langle00|+\frac{1}{6}(|01\rangle_{14}\langle01|+|01\rangle_{14}\langle10|+|10\rangle_{14}\langle01|+|10\rangle_{14}\langle10|)\}
\otimes\nonumber \\
\{\frac{2}{3}|00\rangle_{25}\langle00|+\frac{1}{6}(|01\rangle_{25}\langle01|+|01\rangle_{25}\langle10|+|10\rangle_{25}\langle01|+|10\rangle_{25}\langle10|)\}]+
\nonumber \\
\beta^{2}[\{\frac{2}{3}|00\rangle_{14}\langle00|+\frac{1}{6}(|01\rangle_{14}\langle01|+|01\rangle_{14}\langle10|+|10\rangle_{14}\langle01|+|10\rangle_{14}\langle10|)\}
\otimes\nonumber\\\{\frac{2}{3}|11\rangle_{25}\langle11|+\frac{1}{6}(|01\rangle_{25}\langle01|+|01\rangle_{25}\langle10|+|10\rangle_{25}\langle01|+|10\rangle_{25}\langle10|)\}]+
\nonumber \\
\gamma^{2}[\{\frac{2}{3}|11\rangle_{14}\langle11|+\frac{1}{6}(|01\rangle_{14}\langle01|+|01\rangle_{14}\langle10|+|10\rangle_{14}\langle01|+|10\rangle_{14}\langle10|)\}
\otimes\nonumber \\
\{\frac{2}{3}|00\rangle_{25}\langle00|+\frac{1}{6}(|01\rangle_{25}\langle01|+|01\rangle_{25}\langle10|+|10\rangle_{25}\langle01|+|10\rangle_{25}\langle10|)\}]+
\nonumber \\
\beta\gamma[\{\frac{1}{3}(|00\rangle_{14}\langle01|+|00\rangle_{14}\langle10|)+\frac{1}{3}(|01\rangle_{14}\langle11|+|10\rangle_{14}\langle11|)\}\otimes\nonumber \\
\{\frac{1}{3}(|11\rangle_{25}\langle01|+|11\rangle_{25}\langle10|)+\frac{1}{3}(|01\rangle_{25}\langle00|+|10\rangle_{25}\langle00|)\}]+
\nonumber\\
\gamma\beta[\{\frac{1}{3}(|11\rangle_{14}\langle01|+|11\rangle_{14}\langle10|)+\frac{1}{3}(|01\rangle_{14}\langle00|+|10\rangle_{14}\langle00|)\}
\otimes\nonumber\\\{\frac{1}{3}(|00\rangle_{25}\langle01|+|00\rangle_{25}\langle10|)+\frac{1}{3}(|01\rangle_{25}\langle11|+|10\rangle_{25}\langle11|)\}]
\end{eqnarray}
Now from calculated $\rho_{1245}$,we partially trace out the
particle  4 and 2. The resulting state comes out as,
\begin{eqnarray}
\rho_{15}=\alpha^{2}[\frac{25}{36}|00\rangle_{15}\langle00|+\frac{5}{36}|01\rangle_{15}\langle01|+\frac{5}{36}|10\rangle_{15}\langle10|+\frac{1}{36}|11\rangle_{15}\langle11|]+
\nonumber\\\beta^{2}[\frac{5}{36}|00\rangle_{15}\langle00|+\frac{25}{36}|01\rangle_{15}\langle01|+\frac{1}{36}|10\rangle_{15}\langle10|+\frac{5}{36}|11\rangle_{15}\langle11|]+
\nonumber
\\\gamma^{2}[\frac{25}{36}|10\rangle_{15}\langle10|+\frac{5}{36}|11\rangle_{15}\langle11|+\frac{5}{36}|00\rangle_{15}\langle00|+\frac{1}{36}|01\rangle_{15}\langle01|]+
\nonumber\\\beta\gamma[\frac{4}{9}|01\rangle_{15}\langle10|]+\gamma\beta[\frac{4}{9}|10\rangle_{15}\langle01|]
\end{eqnarray}
Similarly tracing out partially the particle pairs (2,5),(1,4),and
(1,5),we get the following resulting states respectively:
\begin{eqnarray}
\rho_{14}=\alpha^{2}[\frac{2}{3}|00\rangle_{14}\langle00|+\frac{1}{6}(|01\rangle_{14}\langle01|+|01\rangle_{14}\langle10|+|10\rangle_{14}\langle01|+|10\rangle_{14}\langle10|)]+
\nonumber\\\beta^{2}[\frac{2}{3}|00\rangle_{14}\langle00|+\frac{1}{6}(|01\rangle_{14}\langle01|+|01\rangle_{14}\langle10|+|10\rangle_{14}\langle01|+|10\rangle_{14}\langle10|)]+
\nonumber\\\gamma^{2}[\frac{2}{3}|11\rangle_{14}\langle11|+\frac{1}{6}(|01\rangle_{14}\langle01|+|01\rangle_{14}\langle10|+|10\rangle_{14}\langle01|+|10\rangle_{14}\langle10|)]
\end{eqnarray}
\begin{eqnarray}
\rho_{25}=\alpha^{2}[\frac{2}{3}|00\rangle_{25}\langle00|+\frac{1}{6}(|01\rangle_{25}\langle01|+|01\rangle_{25}\langle10|+|10\rangle_{25}\langle01|+|10\rangle_{25}\langle10|)]+
\nonumber\\\beta^{2}[\frac{2}{3}|11\rangle_{25}\langle11|+\frac{1}{6}(|01\rangle_{25}\langle01|+|01\rangle_{25}\langle10|+|10\rangle_{25}\langle01|+|10\rangle_{25}\langle10|)]+
\nonumber\\\gamma^{2}[\frac{2}{3}|00\rangle_{25}\langle00|+\frac{1}{6}(|01\rangle_{25}\langle01|+|01\rangle_{25}\langle10|+|10\rangle_{25}\langle01|+|10\rangle_{25}\langle10|)]
\end{eqnarray}
\begin{eqnarray}
\rho_{42}=\alpha^{2}[\frac{25}{36}|00\rangle_{42}\langle00|+\frac{5}{36}|01\rangle_{42}\langle01|+\frac{5}{36}|10\rangle_{42}\langle10|+\frac{1}{36}|11\rangle_{42}\langle11|]+
\nonumber\\\beta^{2}[\frac{5}{36}|00\rangle_{42}\langle00|+\frac{25}{36}|01\rangle_{42}\langle01|+\frac{1}{36}|10\rangle_{42}\langle10|+\frac{5}{36}|11\rangle_{42}\langle11|]+
\nonumber\\\gamma^{2}[\frac{5}{36}|00\rangle_{42}\langle00|+\frac{1}{36}|01\rangle_{42}\langle01|+\frac{25}{36}|10\rangle_{42}\langle10|+\frac{5}{36}|11\rangle_{42}\langle11|]+
\nonumber\\\beta\gamma[\frac{4}{9}|01\rangle_{42}\langle10|+\frac{4}{9}|10\rangle_{42}\langle01|]
\end{eqnarray}
\\\\
\section{Analysis of the the Separability and Inseparability
condition for the local and non local output states:}First of all
in this section, we obtain the expression of the determinants
$W_3$ and $W_4$ for local and non local output states by using
Peres Horodecki criterion [7,8]. Then we consider the case when
$(\beta= \gamma)$ and investigate the separability and
inseparability criterion for local and nonlocal states. We also
obtained the range of $\alpha^2$ for which the local output states
are
separable and the nonlocal output states are inseparable. \\

We shall now try to find out the expression of $W_3$ and $W_4$ for
the states $\rho_{15}$, $\rho_{14}$,$\rho_{25}$ and $\rho_{42}$,
with the help of Peres-Horodecki
Criterion.\\\\
\textbf{Peres-Horodecki Theorem :}The necessary and sufficient
condition for the state $\rho$ of two spin $\frac{1}{2}$ particles
to be inseparable is that at least one of the eigen values of the
partially transposed operator defined as
$\rho^{T}_{m\mu,n\nu}=\rho_{m\mu,n\nu}$, is negative. This is
equivalent to the condition that at least one of the two
determinants\\
$W_{3}= \begin{array}{|ccc|}
  \rho_{00,00} & \rho_{01,00} & \rho_{00,10} \\
  \rho_{00,01} & \rho_{01,01} & \rho_{00,11} \\
  \rho_{10,00} & \rho_{11,00} & \rho_{10,10}
\end{array}$ and $W_{4}=\begin{array}{|cccc|}
   \rho_{00,00} & \rho_{01,00} & \rho_{00,10} & \rho_{01,10}\\
  \rho_{00,01} & \rho_{01,01} & \rho_{00,11} & \rho_{01,11} \\
  \rho_{10,00} & \rho_{11,00} & \rho_{10,10} & \rho_{11,10} \\
  \rho_{10,01} & \rho_{11,01} & \rho_{10,11} & \rho_{11,11}
\end{array}$\\
is negative.\\
 The values of $W_{3}$ and $W_{4}$ for the different
states $\rho_{15}$,$\rho_{14}$,$ \rho_{25}$ an $\rho_{42}$ are as
follows:\\
\textbf{For non local output states $\rho_{15}=\rho_{42}$ :}\\
$ W_{3}=
\begin{array}{|ccc|}
  \frac{25}{36}\alpha^{2}+\frac{5}{36}\beta^{2}+\frac{5}{36}\gamma^{2} & 0 & 0 \\
  0 & \frac{5}{36}\alpha^{2}+\frac{25}{36}\beta^{2}+\frac{1}{36}\gamma^{2} & 0 \\
  0 & 0 & \frac{5}{36}\alpha^{2}+\frac{1}{36}\beta^{2}+\frac{25}{36}\gamma^{2}
\end{array}$\\\\
 and\\
 $W_{4}=\begin{array}{|cccc|}
   \frac{25}{36}\alpha^{2}+\frac{5}{36}\beta^{2}+\frac{5}{36}\gamma^{2} & 0 & 0 & \frac{4}{9}\beta\gamma\\
   0 & \frac{5}{36}\alpha^{2}+\frac{25}{36}\beta^{2}+\frac{1}{36}\gamma^{2} & 0 & 0 \\
   0 & 0 & \frac{5}{36}\alpha^{2}+\frac{1}{36}\beta^{2}+\frac{25}{36}\gamma^{2} & 0 \\
  \frac{4}{9}\beta\gamma & 0 & 0 & \frac{1}{36}\alpha^{2}+\frac{5}{36}\beta^{2}+\frac{5}{36}\gamma^{2}
\end{array}$\\\\\\

\textbf{For local output states $\rho_{14}$ and $\rho_{25}$ :}\\
\textbf{For $\rho_{14}$:}\\\\
 $W_{3}=
\begin{array}{|ccc|}
  \frac{2}{3}(\alpha^2+\beta^2) & 0 & 0 \\
  0 & \frac{1}{6}& 0 \\
  0 & 0 & \frac{1}{6}
\end{array}$\\\\
and $ W_{4}=
\begin{array}{|cccc|}
  \frac{2}{3}(\alpha^{2}+\beta^{2}) & 0 & 0 & \frac{1}{6} \\
  0 & \frac{1}{6} & 0 & 0 \\
  0 & 0 & \frac{1}{6} & 0 \\
  \frac{1}{6}& 0 & 0 &
  \frac{2}{3}\gamma^2 \\
\end{array}$\\\\

 \textbf{For $\rho_{25}$:}\\\\
  $W_{3}=
\begin{array}{|ccc|}
  \frac{2}{3}(\alpha^2+\gamma^2) & 0 & 0 \\
  0 & \frac{1}{6}& 0 \\
  0 & 0 & \frac{1}{6}
\end{array}$\\\\
and $ W_{4}=
\begin{array}{|cccc|}
  \frac{2}{3}(\alpha^{2}+\gamma^{2}) & 0 & 0 & \frac{1}{6} \\
  0 & \frac{1}{6} & 0 & 0 \\
  0 & 0 & \frac{1}{6} & 0 \\
  \frac{1}{6}& 0 & 0 &
  \frac{2}{3}\beta^2 \\
\end{array}$\\\\

Now if we put $\beta=\gamma$, then the density matrices
representing the subsystems as given by equations (5), (6), (7),
(8) becomes:
\begin{eqnarray}
\rho_{14}=\rho_{25}=\frac{1}{3}(1+\alpha^2)|00\rangle\langle 00
|+\frac{1}{6}(|01\rangle\langle 01|+|10\rangle\langle 10
|\nonumber\\+|01\rangle\langle 10 |+|10\rangle\langle 01|)
+\frac{1}{3}(1-\alpha^2)|11\rangle\langle 11|
\end{eqnarray}
and
\begin{eqnarray}
\rho_{15}=\rho_{42}=\frac{5}{36}(1+4\alpha^2)|00\rangle\langle 00
|+\frac{1}{36}(13-8\alpha^2)(|01\rangle\langle
01|+|10\rangle\langle 10
|)\nonumber\\+\frac{1}{36}(8-8\alpha^2)(|01\rangle\langle 10
|+|10\rangle\langle 01|)
+\frac{1}{36}(5-4\alpha^2)|11\rangle\langle 11|
\end{eqnarray}\\
\textbf{ Separability and Inseparability criterion for the
subsystem $\rho_{14}$ and
$\rho_{25}$ :}\\\\
For these subsystems, $W_3>0$ and $W_4=\frac{1}{6^4}(3-4\alpha^4)$.\\
Now $W_4 \geq 0$ if $\rho_{14}$and $\rho_{25}$ are separable. A
simple calculation reveal that $\rho_{14}$and $\rho_{25}$ are
separable when input parameter $\alpha^2$ lies in the range
(0,.86].\\\\
\textbf{ Separability and Inseparability criterion for the
subsystem $\rho_{15}$ and
$\rho_{42}$:}\\\\
For these subsystems, $W_3>0$ and
$W_4=\frac{1}{36^{4}}(5(1+4\alpha^2)(5-4\alpha^2)-(8-8\alpha^2)^2)$.
For inseparability of $\rho_{15}$and $\rho_{42}$ we must have,
$W_4 \leq 0$. A simple calculation reveal that $\rho_{15}$and
$\rho_{42}$ are inseparable
when input parameter $\alpha^2$ lies in the range (0,.22).\\\\
In the common interval (0, 0.22) for $\alpha^{2}$ although
$\rho_{14}$ and $\rho_{25}$ are  separable, $\rho_{15}$ and
$\rho_{42}$ are entangled.
\section{A comparative study of the Entanglement and Mixed ness of the local and non local output states:}
In this section we have made a comparative study of the mixed ness
and entanglement of the local and non local subsystems when
$\alpha^2\in (0,0.22)$. For that reason first of all we find out
the concurrence and linear entropy of the subsystems to quantify
the amount of entanglement and mixed ness in them respectively. To
find out the amount of entanglement we generally use Wootters
formula of Concurrence.\\
\textbf{Concurrence or Entanglement of Formation:} Wootters [9,10]
gave out, for the mixed state $\hat{\rho}$ of two qubits, the
concurrence is
\begin{eqnarray}
C=max (\lambda_1-\lambda_2-\lambda_3-\lambda_4,0)
\end{eqnarray}
where the $\lambda_i$, in decreasing order, are the square roots
of the eigen values of the matrix $\rho^{\frac{1}{2}}(\sigma_y
 \otimes \sigma_y)\rho^*(\sigma_y \otimes
\sigma_y)\rho^{\frac{1}{2}}$ denotes the complex conjugation of
$\rho$ in the computational basis $\{|00\rangle, |01\rangle,
|10\rangle, |11\rangle\}$ and $\sigma_y$ is the Pauli operator.
The entanglement of formation $E_F$ can then be expressed as a
function of C, namely
\begin{eqnarray}
E_F=-\frac{1+\sqrt{1-C^2}}{2}\log_2\frac{1+\sqrt{1-C^2}}{2}-\frac{1-\sqrt{1-C^2}}{2}\log_2\frac{1-\sqrt{1-C^2}}{2}
\end{eqnarray}\\

 \textbf{ Concurrence for the subsystems
$\rho_{14}$ and
$\rho_{25}$:}\\\\
Here we investigate the amount of entanglement present in these
subsystems when the input parameter $\alpha^2$ lies in the range
(0,.22).\\
$C(\rho_{14},\rho_{25})=2max(\frac{1}{6}-\frac{1}{6}(\sqrt{4-4\alpha^4}),0)=0$
in $0<\alpha^2<0.22$.\\\\
 \textbf{ Concurrence for the subsystems
$\rho_{15}$ and
$\rho_{42}$:}\\\\
Here also we investigate the amount of entanglement present in
these subsystems when the input parameter $\alpha^2$ lies in the
range
(0,.22).\\
$C(\rho_{15},\rho_{42})=2max(\frac{8-8\alpha^2}{36}-\frac{1}{36}(\sqrt{5(1+4\alpha^2)(5-4\alpha^2)}),0)$
Now for $0<\alpha^2<0.22$, the concurrence
$C(\rho_{15},\rho_{42})$ of the subsystems lies in the range
(.001,.17).\\
 \textbf{
Linear Entropy for the subsystems $\rho_{14}$ and
$\rho_{25}$:}\\\\
The expression for the state dependent Linear Entropy  is defined as:\\
$S_{L}(\rho) = \frac{4}{3}(1 - Tr (\rho^{2}))$ [11]\\
 Here we investigate the amount of mixed ness in the subsystems
$\rho_{14}$ and $\rho_{25}$ when the input parameter $\alpha^2$
lies in the range
(0,.22).\\
The linear entropy for these
subsystems is given by,\\
$S_L(\rho_{14})=S_L(\rho_{25})=\frac{8}{27}(3-\alpha^4)$. Now when
$\alpha^2 \in (0,.22)$, then $S_L(\rho_{14},\rho_{25})\in
(.87,.89)$.\\\\
\textbf{ Linear Entropy for the subsystems $\rho_{15}$ and
$\rho_{42}$:}\\\\
The Linear Entropy in this case is: $S_{L}(\rho_{15}) =
S_{L}(\rho_{42}) =
\frac{4}{3}[-\frac{1}{324}(168\alpha^{4}-12\alpha^{2}+129)+1]$\\
 We now investigate the amount
of mixed ness in the subsystems $\rho_{15}$ and $\rho_{42}$ when
the input parameter $\alpha^2$ lies in the range
(0,.22).\\
Now when $\alpha^2 \in (0,.22)$, then $S_L(\rho_{15},\rho_{42})\in
(.77,.81)$.\\\\
From the above calculations of the linear entropy it is clear that
the mixed ness of the non-local outputs are less than the local
output states for the same values of $\alpha^2$. This opens up the
possibility of extracting pure entanglement efficiently from these
two partially entangled state between Alice and Bob.

Since it is evident from the previous section that the local
output states are separable and non local output states are
inseparable when $\alpha^2 \in (0,0.22)$, it remains interesting
to have a comparative study of the mixed ness and entanglement of
the subsystem. Here in the following table we show the mixed ness
and the amount of entanglement of the subsystem when the input
probability
$\alpha^2$ lies in the range (0,0.22).\\\\
{\bf TABLE 2 ($\alpha^2 \in (0,0.22)$):}\\\\
\begin{tabular}{|c|c|c|}
\hline Subsystems & Linear Entropy & Concurrence \\
\hline $\rho_{15}$ and $\rho_{42}$ &
(.77,.81) & (.001,.17)\\
\hline $\rho_{14}$ and
$\rho_{25}$  & (.87,.89)& 0 \\
\hline
\end{tabular}\\\\

\section{Conclusions }
In summary we can say that in this protocol we are able to
generate two bipartite entangled state between two friends Alice
and Bob from a tripartite entangled state initially shared between
Alice, Bob and Carol. We also find out under what conditions the
non-local output states are inseparable and local output states
are separable. We also considered the case when $\beta=\gamma$,
and also made a comparative study between the amount of
entanglement and mixed ness of the local and non local subsystems.
The interesting result that we obtain is that the local outputs ,
although separable , have a higher degree of mixed ness than the
non-local outputs. While comparing the mixed ness and entanglement
of the non-local outputs , we have found out that the mixed ness
and concurrence share a positive correlation. Thus if the mixed
ness of the non-local outputs increases, their amount of
entanglement also increases and consequently we can encode more
information in these non-local outputs. So, we can say that mixed
ness has a positive impact on data encoding.
\section{Acknowledgement}
The authors acknowledge Mr. Satyabrata Adhikari for his rent less
support in completion of this work.
\section{References}
$[1]$ C.H.Bennett, G.Brassard, C.Crepeau, R.Jozsa, A.Peres, W.K.Wootters,
Phys Rev Lett. 70 (1993), 1895\\
$[2]$C.H.Bennett, S.J.Wiesner, Phys Rev Lett, 69 (1992), 2881
\\ $[3]$ P.W.Shor, Phys Rev A 52 (1995) 2493\\
$[4]$ T. Jennewein, C.Simon, G. Weihs, H.Weinfurter and A.Zeilinger, Phys Rev Lett, 84 (200) 4729
\\
$[5]$V.Buzek, V.Vedral, M.B.Plenio, P.L.Knight, M.Hillery, quant-ph/9701028v1,23Jan, 1997\\

$[6]$ V.Buzek, M.Hillery, quant-ph/9607018v1, 20 Jul, 1996.\\
$[7]$M.Horodecki, P.Horodecki, R.Horodecki, PLA 223, 1 (1996)\\
$[8]$A.Peres, PRL 77, 1413, 1996\\
$[9]$ Hill, S and W.K.Wootters,PRL 78, (5022),1997\\
$[10]$ Wootters, W.K, PRL 80, 2245, 1998\\
$[11]$ Quantum Computation and Quantum Information, M.A. Nielsen
and I.L.Chuang, Cambridge University Press.
\end{document}